\documentclass[10pt]{ieeeconf}

\makeatletter
\newcommand*\titleheader[1]{\gdef\@titleheader{#1}}
\AtBeginDocument{%
  \let\st@red@title\@title
  \def\@title{%
    \bgroup\normalfont\footnotesize\vskip-0.9in\@titleheader\par\egroup
    \vskip0.6in\st@red@title}
}
\makeatother

\usepackage{amsmath,amssymb}
\usepackage{graphicx,graphicx,verbatim}
\usepackage{mathrsfs}
\usepackage[dvipsnames]{xcolor}
\usepackage{enumerate}
\usepackage{mathtools}
\usepackage[font=it]{caption}
\usepackage{subcaption}
\usepackage{hyperref}

\graphicspath{{./fig}}

\newtheorem{theorem}{Theorem}

\newtheorem{exx}{Example}
\newtheorem{assumption}{Assumption}

\newenvironment{example}{\begin{exx}}{\hfill \hspace*{1pt} \hfill $\circ$ \end{exx}}

\renewcommand{\Re}{\ensuremath{\mathbb{R}}}

\titleheader{An abridged version of this article will be published in the Proceedings of the 2021 IEEE Conference on Decision and Control (\textcopyright 2021 IEEE).}

\title{Passive soft-reset controllers for nonlinear systems}
\author{Justin H. Le and Andrew R. Teel\thanks{The authors are with the Electrical and Computer Engineering Department, University of California, Santa Barbara, USA. Email: {\em justinle@ucsb.edu}, {\em teel@ucsb.edu}. Research supported in part by the Air Force Office of Scientific Research under grant AFOSR FA9550-18-1-0246.}}

\begin{document}

\maketitle

\begin{abstract}
Soft-reset controllers are introduced as a way to approximate hard-reset controllers. The focus is on implementing reset controllers that are (strictly) passive and on analyzing their interconnection with passive plants. A passive hard-reset controller that has a strongly convex energy function can be approximated as a soft-reset controller. A hard-reset controller is a hybrid system whereas a soft-reset controller corresponds to a differential inclusion, living entirely in the continuous-time domain. This feature may make soft-reset controllers easier to understand and implement. A soft-reset controller contains a parameter that can be adjusted to better approximate the action of the hard-reset controller. Closed-loop asymptotic stability is established for the interconnection of a passive soft-reset controller with a passive plant, under appropriate detectability assumptions. Several examples are used to illustrate the efficacy of soft-reset controllers. 
\end{abstract}

\section{Introduction}

A reset controller is a dynamical system whose state jumps to a new value when a prescribed condition, involving controller states and plant states, is realized.   One of the earliest known reset controllers is due to Clegg, who introduced an integrating circuit that, for all intents and purposes, resets its state to zero when the product of the input and the output of the integrator attempts to become negative \cite{Clegg58}.   The Clegg integrator was extended by Horowitz and co-authors some twenty years later to more general ``first-order reset elements'' (FOREs) \cite{KrishnanHorowitz74}, \cite{HorowitzRosenbaum75}.    
After another two decades, stability and performance results for
linear reset control systems began to receive significant attention, starting with the work of Hollot and co-authors
\cite{Hollot97}, \cite{Hollot-etal97ACC}, \cite{Hollot-etal97CDC}, \cite{Hollot-etal99}, \cite{Hollot-etal2000CDC}, \cite{Hollot-etal04}.   This work was continued and expanded upon within the hybrid systems framework of \cite{Goebel12a} in 
\cite{NesicZaccarianTeel08} and \cite{Zaccarian2011} for example.
Other contributions pertaining to linear reset control systems include \cite{BanosBook12} and \cite{Heemels10}.    See \cite{Zaccarian18} for a recent overview, perspective, and applications of linear reset control systems.   For nonlinear reset control systems, Haddad and co-authors made noteworthy contributions, especially for lossless interconnections \cite{Bupp00a}, \cite{HaddadTAC07} using an alternative hybrid systems framework \cite{HaddadChellaboinaNersesov06b}.      In all of this work, controller resets are typically triggered by the closed-loop system state hitting a surface or attempting to enter a sector.   Other reset control system mechanisms include those found in event-triggered control, where a control signal is typically held constant until a change to its value is required to maintain closed-loop stability or other properties \cite{Astrom02}, \cite{Tabuada07a}.

In this work, we consider an alternative implementation of control systems with resets
triggered by the closed-loop system state attempting to enter a sector.    We give conditions under which such a reset control system can be implemented using a differential inclusion rather than a hybrid system that involves resets.  
(Differential inclusions have appeared in other works related to reset control systems, including \cite{VanLoon16}, \cite{Le2021}, and \cite{Baradaran2021}.) We focus on nonlinear control problems, especially those where passivity plays a role, both in characterizing a nonlinear plant and also in characterizing a reset controller.   Our results are inspired by similar results for linear reset control systems given in \cite{Teel21a}.    In Section \ref{section:hard-reset-controllers}, we specify the passive, hard-reset controllers that we aim to implement as soft-reset controllers.  We emphasize that such passive hard-reset controllers should admit a strongly convex energy function in order to be
implementable as a soft-reset controller.  Our passive soft-reset controllers are introduced in Section \ref{section:soft-reset-controllers}.    In Section \ref{section:interconnections}, we discuss the interconnection of a passive soft-reset controller with a passive plant.  Section \ref{section:Illustrations} contains several illustrations of the developed theory.

\section{Notation}
For $x \in \Re^{n}$, we use
$|x|:= \sqrt{x^{T} x}$.     Given a pair $(x,u) \in \Re^{n} \times \Re^{m}$, by abuse
of notation we sometimes consider $z:=(x,u)$ to be  a vector in $\Re^{n+m}$.  By the same
abuse of notation, we sometimes write a function defined on $\Re^{n} \times \Re^{m}$ as
a function defined on $\Re^{n + m}$, i.e.,  $f(x,u)$ may be written as $f(z)$ with $z=(x,u)$.
A function $f:\Re^{n} \rightarrow \Re^{m}$ is said to be {\em sector bounded near the origin} if there exist
$\delta>0$ and $L>0$ such that $|f(z)| \leq L |z|$ for all $z \in \Re^{n}$ satisfying $|z| \leq \delta$.
It is said to be {\em quadratically bounded near the origin} if there exist $\delta>0$ and $L>0$
such that $|f(z)| \leq L |z|^{2}$ for all $z \in \Re^{n}$ satisfying $|z| \leq \delta$.
A function $V:\Re^{n} \rightarrow \Re$ is said to be {\em convex} if 
\begin{align*}
V\left( \lambda x_{1} + (1-\lambda) x_{2} \right) \leq \lambda V(x_{1}) + (1-\lambda) V(x_{2})
\end{align*}
 for all $x_{1},x_{2} \in \Re^{n}$ and all $\lambda \in \left[ 0,1 \right]$.  
It is said to be {\em strongly convex} if there exists $\mu>0$ such that $x \mapsto V(x) - \mu |x|^{2}$ is convex.
 A continuously differentiable function $V:\Re^{n} \rightarrow \Re$
is strongly convex if and only if there exists $\mu>0$ such that, for all
$x,y \in \Re^{n}$,
\begin{align}
\label{eq:strongly-convex}
V(y) \geq V(x) + \langle \nabla V(x) , y-x \rangle + \mu | x-y|^{2} .
\end{align}

\section{Passive hard-reset controllers}
\label{section:hard-reset-controllers}

A (square) hard-reset control system is a hybrid system with state $x_{c} \in \Re^{n_{c}}$, input $u_{c} \in \Re^{m_{c}}$,
and output $y_{c} \in \Re^{m_{c}}$
with the following model:
\begin{subequations}
\label{eq:1}
\begin{align}
  (x_{c},u_{c})  & \in C    \qquad \dot{x}_{c} = f_{c}(x_{c},u_{c}) \label{eq:flows} \\
  (x_{c},u_{c}) & \in D   \qquad x_{c}^{+} = g_{c}(x_{c},u_{c}) \label{eq:jumps} \\
                       &     y_{c} = h_{c}(x_{c},u_{c})  \label{eq:output}
\end{align}
\end{subequations}
where, typically, 
\begin{subequations}
\label{eq:2}
\begin{align}
C & :=\left\{ (x_{c},u_{c}) \in \Re^{n_{c}} \times \Re^{m_{c}} :  \varphi(x_{c},u_{c}) \leq 0 \right\}  \label{eq:flow-set} \\
D & := \left\{ (x_{c},u_{c}) \in \Re^{n_{c}} \times \Re^{m_{c}} : \varphi(x_{c},u_{c}) \geq 0 \right\} \label{eq:jump-set} .
\end{align}
\end{subequations}
We impose the following assumptions on the functions that prescribe the model.
\begin{assumption}
\label{assume:big}
The following conditions hold for
the functions $f_{c}, g_{c}, h_{c}$, and $\varphi$ that appear in (\ref{eq:1})-(\ref{eq:2}):
\begin{enumerate}
\item 
\label{assume:1}
({\em Continuous, locally sector bounded data and quadratic jump condition}) \\
There exists $M=M^{T}$ such that
$\varphi(z_{c}) = z_{c}^{T} M z_{c}$ for all $z_{c} \in \Re^{n_{c}+m_{c}}$; also,
$f_c$, $g_c$, and $h_c$ are sector bounded near the origin and continuous.
\item
\label{assume:3}
({\em Jumps land in flow set}) \\
$(x_{c},u_{c}) \in D$ implies $(g_{c}(x_{c},u_{c}),u_{c}) \in C$.
\item
\label{assume:2}
{\em (Passivity via a strongly convex energy function}) \\
There exist a strongly convex, positive definite, continuously differentiable function $V_{c}:\Re^{n_{c}} \rightarrow \Re_{\geq 0}$, a continuous, positive definite, quadratically bounded near the origin function $\rho:\Re^{m_{c}} \rightarrow \Re_{\geq 0}$, and $\varepsilon>0$ such that, with the definition
\begin{align}
C_{\varepsilon}:= \left\{ z_{c} \in \Re^{n_{c}+m_{c}} : z_{c}^{T} M z_{c} \leq \varepsilon z_{c}^{T} z_{c} \right\} ,
\label{eq:Ceps}
\end{align}
we have
\begin{subequations}
\begin{align}
\langle \nabla V_{c}(x_{c}), \! f_{c}(x_{c},u_{c}) \rangle & \leq y_{c}^{T} u_{c} \! - \! \rho(y_{c})   \ \ \forall (x_{c},u_{c}) \in C_{\varepsilon} \label{eq:energy-flows} \\
  V_{c}(g_{c}(x_{c},u_{c})) & \leq V_{c}(x_{c}) \qquad \quad \forall (x_{c},u_{c}) \in D 
  \label{eq:energy-jumps}
\end{align}
\end{subequations}
where $y_c=h_{c}(x_{c},u_{c})$.
\item
\label{assume:4}
({\em Minimum phase and detectable}) \\
Any absolutely continuous, bounded solution $(x_{c},u_{c}):[0,\infty) \rightarrow \Re^{n_{c}} \times \Re^{m_{c}}$ 
of (\ref{eq:flows}), i.e., for almost all $t \in [0,\infty)$,
\begin{align}
(x_{c}(t),u_{c}(t)) \in C  ,  \quad \dot{x}_{c}(t) = f_{c}(x_{c}(t),u_{c}(t)) ,
\end{align}
that satisfies $y_{c}(t) = 0$ for all $t \in [0,\infty)$ also satisfies
$\lim_{t \rightarrow \infty} u_{c}(t) =0$ and $\lim_{t \rightarrow \infty} x_{c}(t) = 0$.
\hfill  $\blacksquare$
\end{enumerate}
\end{assumption}

\vspace{.1in}

Assumption \ref{assume:big}.\ref{assume:3} implies
that, after a jump, the solution of the hard-reset system (\ref{eq:1})-(\ref{eq:2}) has the potential to flow without immediately jumping again.  However, it is possible that $(g_{c}(x_{c},u_{c}),u_{c}) \in C \cap D$, in which case the solution also has the potential to jump immediately.  That is, there is no guarantee that all of the complete solutions of the hard-reset control system (\ref{eq:1})-(\ref{eq:2}) have time domains that are unbounded in the ordinary time direction. This is one of the primary motivations for considering a ``soft'' implementation of the reset control system (\ref{eq:1})-(\ref{eq:2}), as we do in the next section.


The crux of Assumption \ref{assume:big} is Assumption \ref{assume:big}.\ref{assume:2}, which imposes
a type of strict passivity condition on the hard-reset control system (\ref{eq:1})-(\ref{eq:2}).  For more context, compare with
Assumption \ref{assume:plant}.\ref{assume:p2}, where strict passivity of a continuous-time, passive plant
is characterized. In addition, we use (\ref{eq:energy-jumps}) and the strong convexity of $V_{c}$ in the next section to propose the soft-reset implementation of the hard-reset system \eqref{eq:1}.

\vspace{.05in}

The following example illustrates a class of systems satisfying Assumption \ref{assume:big}.

\begin{example}
\label{example}
Consider a hard-reset control system (\ref{eq:1})-(\ref{eq:2}) with $(x_{c},u_{c}) \in \Re \times \Re$,
the following data:
\begin{subequations}
\label{eq:example-data}
\begin{align}
 f_{c}(x_{c},u_{c}) & : =  a_{c} x_{c} + b_{c} u_{c} \\
 g_{c}(x_{c},u_{c}) & : = r_{c} x_{c} +  p_{c} u_{c} \\
 h_{c}(x_{c},u_{c}) & : =  c_{c} x_{c}  + d_{c} u_{c} \\
  M & := \left[ \begin{array}{cc} m_{11} & m_{12} \\ m_{12} & m_{22} \end{array} \right]
\end{align}
\end{subequations}
and the energy function
\begin{align}
\label{eq:example-energy}
    V_{c}(x_{c}) : = \kappa x_{c}^{2}
\end{align}
where $\kappa>0$.  There are no assumptions yet on the signs of the parameters in (\ref{eq:example-data}). Define
\begin{align}
\label{eq:M0}
 M_{0} : = \left[ \begin{array}{cc} 2 \kappa a_{c} + \rho c_{c}^{2}  & \kappa b_{c} + \rho c_{c} d_{c}  - 0.5 c_{c} \\
 \kappa b_{c} +\rho c_{c} d_{c}- 0.5 c_{c} & -d_{c}  + \rho d_c^{2} \end{array} \right]
\end{align}
and note that if 
\begin{align}
\left[ \begin{array}{c} x_{c} \\ u_{c} \end{array} \right]^{T} M_{0} \left[ \begin{array}{c} x_{c} \\ u_{c} \end{array} \right] \leq 0
\end{align}
then
\begin{align}
   \langle \nabla V_{c}(x_{c}) , f_{c}(x_{c},u_{c}) \rangle & = 2 \kappa a_{c} x_{c}^{2} + 2 \kappa b_{c} x_{c} u_{c} 
   \nonumber \\
  &  \leq  (c_{c} x_{c} \! + \! d_{c} u_{c}) u_{c} \! - \! \rho (c_{c} x_{c} \! + \! d_{c} u_{c})^{2} \nonumber \\
  & = y_{c} u_{c} - \rho  y_{c}^{2} .  \label{eq:14}
\end{align}
Let $\varepsilon>0$.
By the S-procedure \cite[p. 655]{BoydVandenberghe},
if there exists $\lambda \in [0,\infty)$
such that
\begin{align}
\label{eq:S-procedure}
\lambda (M - \varepsilon I) - M_{0} \geq 0
\end{align}
and there exists $z_{c} \neq 0$ such that
$z_{c}^{T} M z_{c} \leq 0$, i.e., the flow set $C$ contains more points than just the origin,
 then (\ref{eq:14}) holds for $z_{c} \in C_{\varepsilon}$ where $C_{\varepsilon}$ is defined in
(\ref{eq:Ceps}).    For example, when $d_{c}>0$, we can take 
$M=M_{0} + \varepsilon I$, with $\varepsilon>0$ and $\rho>0$ small, and $\lambda =1$.
In this case, there exists $z_{c} \neq 0$ of the form $z_{c}=(0,u_{c})$ with $u_{c} \neq 0$ such that
$z_{c}^{T} M z_{c} \leq 0$ so that the S-procedure applies.  It also follows that, with the zero reset map, i.e.,
$r_{c}=p_{c}=0$, $(x_{c},u_{c}) \in D$ implies $(g(x_{c},u_{c}),u_{c}) \in C$, i.e., Assumption \ref{assume:big}.\ref{assume:3} holds.  If $d_{c}>0$
and $b_{c} c_{c} > a_{c} d_{c}$ then the minimum phase and detectability properties hold.
\end{example}

\section{Passive soft-reset controllers}
\label{section:soft-reset-controllers}
Using (\ref{eq:strongly-convex}),
the strong convexity assumption on $V_{c}$ in Assumption \ref{assume:big}.\ref{assume:2} and the bound in (\ref{eq:energy-jumps}) imply that there exists $\mu>0$ such that
\begin{align}
& (x_{c},u_{c}) \in D \quad \Longrightarrow \qquad \nonumber \\
& \langle \nabla V_{c}(x_{c}) , x_{c} - g_{c}(x_{c},u_{c}) \rangle \geq \mu | x_{c} - g_{c}(x_{c},u_{c})|^{2} .
\end{align}
With (\ref{eq:jump-set}),
another way to write this condition is as
\begin{align}
& s \in \mbox{\rm SGN}(\varphi(x_{c},u_{c})) \Longrightarrow \nonumber \\
& \qquad \langle \nabla V_{c}(x_{c}), (s+1) (g_{c}(x_{c},u_{c}) - x_{c}) \rangle \nonumber \\
& \qquad \qquad \qquad \qquad  \leq - (s+1) \mu  | x_{c} - g_{c}(x_{c},u_{c})|^{2} , \label{eq:7}
\end{align}
where
the set-valued mapping $\mbox{\rm SGN}:\Re \rightrightarrows \Re$ is defined as
$\mbox{\rm SGN}(s) := \mbox{\rm sign}(s)$ for $s \neq 0$ and
$\mbox{\rm SGN}(0) := [-1,1]$.
Thus, we can add a term of the form
\begin{align}
\label{eq:add-this}
    - \gamma(x_{c},u_{c}) 
\left( 
\vphantom{\frac{1}{2}} 
\mbox{\rm SGN} 
\left( 
\vphantom{\frac{}{}} 
\varphi(x_{c},u_{c}) 
\right) 
+ 1 
\right) 
\left( 
\vphantom{\frac{}{}} x_{c} - g_{c}(x_{c},u_{c}) 
\right)
\end{align}
to the differential equation $\dot{x}_{c} = f_{c}(x_{c},u_{c})$ without increasing
the directional derivative of $V_{c}$ as long as $\gamma$ takes nonnegative values.
Moreover, due to Assumption \ref{assume:big}.\ref{assume:1} and Assumption \ref{assume:big}.\ref{assume:3},  it can be
shown that $|g_{c}(x_{c},u_{c}) - x_{c} |>0$ when $\varphi(x_{c},u_{c})>0$, which means that the additional
term
(\ref{eq:add-this}) can be used to enhance the negativity of the directional derivative of $V_{c}$ outside of the
set $C_{\varepsilon}$ defined in (\ref{eq:Ceps}), potentially obviating the need for
hard resets.

Led by these observations,
the soft-reset implementation of the hard-reset system (\ref{eq:1}) is given by the differential
inclusion
\begin{align}
\label{eq:soft}
\dot{x}_{c} & \in f_{c}(x_{c},u_{c}) \\
  &  - \gamma(x_{c},u_{c}) \left( \vphantom{\frac{1}{2}} \mbox{\rm SGN} \left( \vphantom{\frac{}{}} \varphi(x_{c},u_{c}) \right) + 1 \right) \left( \vphantom{\frac{}{}} x_{c} - g_{c}(x_{c},u_{c}) \right)  \nonumber
\end{align}
where $\gamma:\Re^{n_{c}} \times \Re^{m_{c}} \rightarrow \Re_{>0}$ is continuous and the set-valued mapping $\mbox{\rm SGN}:\Re \rightrightarrows \Re$ is defined above \eqref{eq:add-this}. Notice that, within the set $C$ defined in \eqref{eq:flows}, the soft-reset dynamics \eqref{eq:soft} match the dynamics \eqref{eq:flows} of the hard-reset controller. Outside of $C$, the dynamics \eqref{eq:soft} imitate a reset (if $\gamma(x_{c},u_{c})$ is large) by rapidly driving $x_{c}$ towards $g_{c}(x_{c},u_{c})$.

We now show that the soft-reset control system (\ref{eq:soft}) inherits a strict passivity property from
its hard-reset control system inspiration (\ref{eq:1})-(\ref{eq:2}) when $\gamma$ is sufficiently large. In addition, we note that $\gamma$ does not need to be large when the inequality in (\ref{eq:energy-flows}) holds 
for all $(x_{c},u_{c}) \in \Re^{n_{c}} \times \Re^{m_{c}}$, i.e., $\varepsilon \geq \overline{\sigma}(M)$ where $\overline{\sigma}(M)$ the maximum singular value of $M$.

\begin{theorem}
\label{theorem:1}
If Assumption \ref{assume:big}.\ref{assume:1}, \ref{assume:big}.\ref{assume:3}, and \ref{assume:big}.\ref{assume:2} hold then there exists a continuous, positive definite function
$\sigma_{1}:\Re^{n_{c}} \times \Re^{m_{c}} \rightarrow \Re_{\geq 0}$ and for each $\gamma_{0} >0$ there exists a continuous
function $\gamma:\Re^{n_{c} \times m_{c}} \rightarrow \Re_{>0}$ such that, for each
$(x_{c},u_{c}) \in \Re^{n_{c}} \times \Re^{m_{c}}$ and $s \in \mbox{\rm SGN}(\varphi(x_{c},u_{c}))$,
\begin{align}
& \langle V_{c}(x_{c}), f_{c}(x_{c},u_{c})  - \gamma(x_{c},u_{c})  (s+1) (x_{c} - g_{c}(x_{c},u_{c}) ) \rangle
 \nonumber \\
 & \quad \leq y_{c}^{T} u_{c} - \rho(y_{c})  \label{eq:passive-soft-reset} \\
 & \qquad \qquad  - \gamma_{0} \max \left\{ 0 , \varphi(x_{c},u_{c}) \right\} \frac{\varphi(x_{c},u_{c})}{\max\left\{1,\sigma_{1}(x_{c},u_{c})^{2} \right\}} .  \nonumber
 \end{align}
 One can take $\gamma$ proportional to $\gamma_{0}$ in the case where the inequality in (\ref{eq:energy-flows}) holds for all $(x_{c},u_{c}) \in \Re^{n_{c}} \times \Re^{m_{c}}$.
\end{theorem}

\vspace{.05in}

{\em Proof.}
Define $\widehat{g}_{c}(z_{c}):=\left[ \begin{array}{cc} g_{c}(x_{c},u_{c})^{T} & u_{c}^{T} \end{array} \right]^{T}$. 
Due to Assumption \ref{assume:big}.\ref{assume:1}, which supposes that $g_{c}$ is
sector bounded near the origin and continuous, there exists a continuous function
$\sigma_{1}:D \rightarrow \Re_{\geq 0}$ that is sector bounded near the origin and positive definite
satisfying
\begin{align}
 | M  (\widehat{g}_{c}(z_{c}) + z_{c}) | \leq \sigma_{1}(z_{c})   \qquad \forall z_{c} \in D . 
 \end{align}
 Then using the Cauchy-Schwarz inequality, $M=M^{T}$, and Assumption \ref{assume:big}.\ref{assume:3}, which gives that $\widehat{g}_{c}(z_{c})^{T} M \widehat{g}_{c}(z_{c}) \leq 0$ when
 $z_{c}^{T} M z_{c} \geq 0$, it follows that
 \begin{align}
 z_{c} \neq 0, \ z_{c}^{T} M z_{c}  \geq 0 
  \Longrightarrow &  \nonumber \\
 | g_{c}(x_{c},u_{c}) - x_{c}|_{2} & \geq \frac{-(\widehat{g}(z_{c}) - z_{c})^{T} M ( \widehat{g}(z_{c})+ z_{c})}{\sigma_{1}(z_{c})} \nonumber \\
 & = \frac{ z_{c}^{T} M z_{c} - \widehat{g}(z_{c})^{T} M \widehat{g}(z_{c})}{\sigma_{1}(z_{c})} \label{eq:9} \\
 & \geq  \frac{ z_{c}^{T} M z_{c}}{\sigma_{1}(z_{c})} .  \nonumber
 \end{align}
 Combining (\ref{eq:7}), (\ref{eq:9}), and Assumption \ref{assume:big}.\ref{assume:1} gives
 \begin{align}
 & z_{c} \neq 0 , s \in \mbox{\rm SGN}(z_{c}^{T} M z_{c}) \Longrightarrow \nonumber \\
&  \qquad \langle \nabla V_{c}(x_{c}), (s+1)(g_{c}(x_{c},u_{c}) - x_{c}) \rangle \label{eq:after-9} \\
& \qquad \qquad \qquad \leq - 2 \mu \max \left\{ 0 , z_{c}^{T} M z_{c}
 \right\}  \frac{ z_{c}^{T} M z_{c}}{\sigma_{1}(z_{c})^{2}} . \nonumber
 \end{align}
 Next,
 due to Assumption \ref{assume:big}.\ref{assume:1}, which supposes that $f_{c}$ and $h_{c}$ are sector bounded near the origin and continuous, and due to Assumption \ref{assume:big}.\ref{assume:2} which supposes that $V_{c}$ is continuously differentiable
 and zero at zero, so that $\nabla V_{c}$ is sector bounded near the origin and continuous, and also supposes that $\rho$ is quadratically bounded near the origin, there exists a continuous function
$\sigma_{2}:\Re^{n_{c} + m_{c}} \rightarrow \Re_{\geq 0}$ that is quadratically bounded near the origin such that,
for all $(x_{c},u_{c}) \in \Re^{n_{c}+m_{c}}$,
 \begin{align}
  \langle \nabla V_{c}(x_{c}), f_{c}(x_{c},u_{c}) \rangle - y_{c}^{T} u_{c} + \rho(y_{c})
  \leq \sigma_{2}(z_{c}) .
  \label{eq:12}
 \end{align}
 Note that, according to (\ref{eq:energy-flows}), we can take $\sigma_{2}(z_{c})=0$ for all $z_{c} \in C_{\varepsilon}$. Also note that if $\varepsilon \geq \overline{\sigma}(M)$, the latter denoting the maximum singular value of $M$, then $C_{\varepsilon} = \Re^{n_{c}} \times \Re^{m_{c}}$.
 It follows from (\ref{eq:12}) that
 \begin{align}
 & z_{c} \neq 0, z_{c}^{T} M z_{c} \geq \varepsilon |z_{c}|^{2}  \Longrightarrow \nonumber \\
 &  \langle \nabla V_{c}(x_{c}), f_{c}(x_{c},u_{c}) \rangle - y_{c}^{T} u_{c} + \rho(y_{c})
  \leq \sigma_{2}(z_{c}) \nonumber \\
 & \leq \sigma_{2}(z_{c}) \max \left\{0 , z_{c}^{T} M z_{c} \right\} \frac{z_{c}^{T} M z_{c}}{ \varepsilon^{2} |z_{c}|^{4}} \nonumber \\
&  = \frac{\sigma_{1}(z_{c})^{2} \sigma_{2}(z_{c})}{\varepsilon^{2} |z_{c}|^{4}} \max \left\{0, z_c^{T} M z_{c} \right\}
  \frac{z_{c}^{T} M z_{c}}{\sigma_{1}(z_{c})^{2}}  .
  \label{eq:combine-with-after-9}
 \end{align}
 We note that, since $\sigma_{1}$ is sector bounded near the origin and $\sigma_{2}$
 is quadratically bounded near the origin,
 \begin{align}
 \limsup_{z_{c} \rightarrow 0, z_{c} \in D \backslash \left\{ 0 \right\}} \frac{\sigma_{1}(z_{c})^{2} \sigma_{2}(z_{c})}{ |z_{c}|^{4}} < \infty .
 \end{align}
 Pick $\gamma:\Re^{n_{c}+m_{c}} \rightarrow \Re_{\geq 0}$ to be a continuous function such that, for all $z_{c}  \in D \backslash \left\{ 0 \right\}$,
 \begin{align}
 \gamma(z_{c}) \geq \frac{1}{2\mu} \left( \gamma_{0}  + \frac{\sigma_{1}(z_{c})^{2} \sigma_{2}(z_{c})}{ \varepsilon^{2}|z_{c}|^{4}} \right) .
 \label{eq:24bb}
 \end{align}
 It then follows by combining (\ref{eq:after-9}), (\ref{eq:combine-with-after-9}), and (\ref{eq:24bb})  that
 \begin{align}
 & z_{c} \neq 0 , s \in \mbox{\rm SGN}\left( z_{c}^{T} M z_{c} \right) \Longrightarrow \nonumber \\
 & \langle V_{c}(x_{c}), f_{c}(x_{c},u_{c})  - \gamma(x_{c},u_{c})  (s+1) (x_{c} - g_{c}(x_{c},u_{c}) ) \rangle
 \nonumber \\
 & \leq y_{c}^{T} u_{c} - \rho(y_{c}) - \gamma_{0} \max \left\{ 0 , z_{c}^{T} M z_{c} \right\} \frac{z_{c}^{T} M z_{c}}{\sigma_{1}(z_{c})^{2}}
 \end{align}
 and thus
 \begin{align}
 & s \in \mbox{\rm SGN}\left( z_{c}^{T} M z_{c} \right) \Longrightarrow \nonumber \\
 & \langle V_{c}(x_{c}), f_{c}(x_{c},u_{c})  - \gamma(x_{c},u_{c})  (s+1) (x_{c} - g_{c}(x_{c},u_{c}) ) \rangle
 \nonumber \\
 & \leq y_{c}^{T} u_{c} - \rho(y_{c}) - \gamma_{0} \max \left\{ 0 , z_{c}^{T} M z_{c} \right\} \frac{z_{c}^{T} M z_{c}}{\max\left\{1,\sigma_{1}(z_{c})^{2} \right\}} .
 \end{align}
 Since $\varphi(x_{c},u_{c})=z_{c}^{T} M z_{c}$,
this bound gives the result.
\null \hfill \null $\blacksquare$

\section{Negative feedback interconnection with a passive plant}
\label{section:interconnections}

In this section, we consider the negative feedback interconnection of (\ref{eq:soft}) with a continuous-time,
passive, detectable plant that has state $x_{p} \in \Re^{n_{p}}$, input $u_{p} \in \Re^{m_{p}}$, and
output $y_{p} \in \Re^{m_{p}}$, and can be modeled as
\begin{align}
\label{eq:plant}
\dot{x}_{p} = f_{p}(x_{p},u_{p})     \qquad y_{p}=h_{p}(x_{p})  .
\end{align}
A negative feedback interconnection is produced via the conditions
\begin{subequations}
\label{eq:negative-feedback}
\begin{align}
   u_{c}  = - y_{p} & = - h_{p}(x_{p})  \\
   u_{p}  = y_{c} & = h(x_{c},u_{c}) = h(x_{c},-h_{p}(x_{p})) .
\end{align}
\end{subequations}
The resulting closed-loop system has state variable denoted $x \coloneqq (x_{p}, x_{c})$.
\begin{assumption}
\label{assume:plant}
The following conditions hold for the functions $f_{p}$ and $h_{p}$ that appear in (\ref{eq:plant}):
\begin{enumerate}
\item
\label{assume:p1}
({\em Continuous data}) \\
The functions $f_{p}$ and $h_{p}$ are continuous.
\item
\label{assume:p2}
({\em Passive dynamics}) \\
There exists a continuously differentiable, positive definite, radially unbounded function $V_{p}:\Re^{n_{p}} \rightarrow \Re_{\geq 0}$ such that,
for all  $(x_{p},u_{p}) \in
\Re^{n_{p}} \times \Re^{m_{p}}$,
\begin{align}
\label{eq:passive-plant}
\langle \nabla V_{p}(x_{p}) , f_{p}(x_{p},u_{p}) \rangle \leq y_{p}^{T} u_{p}
\end{align}
where $y_{p} = h_{p}(x_{p})$.
\item
\label{assume:p3}
({\em Detectability}) \\
Each bounded solution $x_{p}:[0,\infty) \rightarrow \Re^{n_{p}}$
of (\ref{eq:plant}) that satisfies $u_{p}(t) = y_{p}(t)=0$ for all $t \in [0,\infty)$
also satisfies $\lim_{t \rightarrow \infty} x_{p}(t) = 0$.
\null \hfill \null $\blacksquare$
\end{enumerate}
\end{assumption}

\vspace{.05in}

\begin{theorem}
\label{thm:feedback}
Suppose Assumption \ref{assume:big} and Assumption \ref{assume:plant} hold.
Then the interconnection of the plant (\ref{eq:plant}) with the soft-reset
system (\ref{eq:soft}) via negative feedback (\ref{eq:negative-feedback}) has the origin
globally asymptotically stable whenever the function $\gamma: \Re^{n_{c} + m_{c}} \rightarrow \Re_{>0}$
in (\ref{eq:soft})  induces the inequality (\ref{eq:passive-soft-reset}).
\end{theorem}

\vspace{.05in}

{\em Proof.}
We use $x:=(x_{p},x_{c})$ for the composite state and $F(x)$ to denote the right-hand side of the closed-loop differential inclusion, so that $\dot{x} \in F(x)$.

Consider the composite Lyapunov function candidate
\begin{align}
V(x):= V_{p}(x_{p}) + V_{c}(x_{c})    \quad \forall (x_{p},x_{c}) \in \Re^{n_{p}} \times \Re^{n_{c}} 
\end{align}
where $V_{p}$ comes from Assumption \ref{assume:plant}.\ref{assume:p2} and $V_{c}$ comes
from Assumption \ref{assume:big}.\ref{assume:2}.
This function is continuously differentiable and positive definite by these assumptions.   It is radially unbounded
since $V_{p}$ is assumed to be radially unbounded and $V_{c}$ is assumed to be strongly convex.
Using (\ref{eq:passive-plant}) and (\ref{eq:passive-soft-reset}) together with (\ref{eq:negative-feedback}), we get, for all $x \in \Re^{n_{p}+n_{c}}$ and all $f \in F(x)$,
\begin{align}
& \langle \nabla V(x) , f \rangle \leq - \rho(y_{c})  \label{eq:25} \\
& \qquad   - \gamma_{0} \max \left\{ 0 , \varphi(x_{c},-y_{p}) \right\} \frac{\varphi(x_{c},-y_{p})}{\max\left\{1,\sigma_{1}(x_{c},-y_{p})^{2} \right\}}    \nonumber
\end{align}
where $\gamma_{0}>0$, $\rho$ is positive definite, and $y_{p}=h_{p}(x_{p})$.  It follows that the right-hand side of (\ref{eq:25})
is never positive and so
the origin is stable and all solutions are bounded.  

To establish global convergence to the origin, we use the invariance principle for differential inclusions
\cite{Ryan91,Ryan98}, which applies due to
Assumption \ref{assume:big}.\ref{assume:1} and \ref{assume:plant}.\ref{assume:p1} and the fact that
the set-valued mapping $\mbox{\rm SGN}$ is outer semicontinuous and bounded with nonempty convex values.
According to the invariance principle, every solution converges to the origin if and only if there does not exist
a solution $x:[0,\infty) \rightarrow \Re^{n}$ and constant $c>0$ satisfying $V(x(t))=c$ for all $t \in [0,\infty)$. 
To rule out such a solution,
we decompose $F$ so that
\begin{align}
\label{eq:closed-loop}
\dot{x} \in F(x)=:g(x)+  \left( \vphantom{\frac{}{}} \mbox{\rm SGN}(\phi(x)) + 1 \right)  \omega(x)
\end{align}
for appropriate, continuous functions $g$, $\phi$ and $\omega$ and note that
\begin{align}
\label{eq:bound-omega}
|\omega(x)| = \gamma(x_{c},-h_{p}(x_{p})) | x_{c} - g_{c}(x_{c},-h_{p}(x_{p})) | 
\end{align}
and
\begin{align}
& \langle \nabla V(x), \omega(x) \rangle \label{eq:need-this} \\
& = - \gamma \left( \vphantom{\frac{}{}} x_{c},-h_{p}(x_{p}) \right) \langle \nabla V_{c}(x_{c}), \left( 
\vphantom{\frac{}{}} x_{c} - g_{c}(x_{c},-h_{p}(x_{p})) \right) \rangle  .
\nonumber
\end{align}
Now suppose there exists a
solution $x:[0,\infty) \rightarrow \Re^{n}$ and constant $c>0$ satisfying $V(x(t))=c$ for all $t \in [0,\infty)$. 
Being a solution of (\ref{eq:closed-loop}), $x(\cdot)$ satisfies, for almost all $t \in [0,\infty)$,
\begin{subequations}
\label{eq:is-a-solution}
\begin{align}
\dot{x}(t) & = g(x(t)) + \left( \vphantom{\frac{}{}} s(t)+1 \right) \omega(x(t)) \\
 s(t) & \in \mbox{\rm SGN}\left( \vphantom{\frac{}{}} \phi(x(t)) \right) .
\end{align}
\end{subequations}
Since $\rho$ is positive definite and $\gamma_{0}>0$, it follows from (\ref{eq:25}) that
such a solution requires $y_{c}(t)=u_{p}(t) = 0$ for all $t \in [0,\infty)$
and $\varphi(x_{c}(t),-y_{p}(t))  \leq 0$ for all $t \in [0,\infty)$, i.e., for almost all $t \geq 0$,
\begin{align}
\label{eq:constraint}
 (x_{c}(t),u_{c}(t)) \in C .
\end{align}
In turn, it follows from 
(\ref{eq:energy-flows}), (\ref{eq:passive-plant}),
(\ref{eq:need-this}), (\ref{eq:7}) and the positivity of $\gamma(\cdot)$ that, for almost all $t \geq 0$,
\begin{subequations}
\begin{align}
0 & = \langle \nabla V(x(t)), g(x(t)) \rangle \\
0 &  = \langle \nabla V(x(t)), (s(t) + 1) \omega(x(t)) \rangle .
\end{align}
\end{subequations}
Again with (\ref{eq:7}), the positivity of $\mu$ and the strict positivity of $\gamma(\cdot)$, it follows
that, for almost all $t \geq 0$,
\begin{align}
 (s(t)+1)  | x_{c} - g_{c}(x_{c},-y_{p}(t)) |^{2} = 0 
\end{align}
and thus, from (\ref{eq:bound-omega}), for almost all $t \geq 0$,
\begin{align}
 (s(t)+1) \omega(x(t)) = 0 .
\end{align}
It then follows from (\ref{eq:is-a-solution}) that $x(\cdot)$ is a solution of $\dot{x}=g(x)$, with $g$ defined
via (\ref{eq:closed-loop}).  In particular, with (\ref{eq:constraint}), $x_{c}(\cdot)$ is a solution of (\ref{eq:flows}).   Then, according to Assumption \ref{assume:big}.\ref{assume:4}, it follows that $\lim_{t \rightarrow \infty} u_{c}(t) =\lim_{t \rightarrow \infty} - y_{p}(t) = 0$ and 
$\lim_{t \rightarrow \infty} x_{c}(t)=0$.   We thus have $\lim_{t \rightarrow \infty} V_{p}(x_{p}(t))=c$.  Again, by the invariance principle, there must exist a solution $\hat{x}_{p}:[0,\infty) \rightarrow \Re^{n}$
 of (\ref{eq:plant}) with $V_{p}(\hat{x}_{p}(t)) =c$ and $\hat{u}_{p}(t)=\hat{y}_{p}(t)=0$ for all $t \geq 0$.
 However, the existence of such a solution contradicts Assumption \ref{assume:plant}.\ref{assume:p3}.
 Hence, there is no solution that keeps $V$ equal to a non-zero constant.  Thus, the origin is globally attractive and global asymptotic stability is established.
\null \hfill \null $\blacksquare$

While our assumptions focus on guaranteeing global results, it is clear that local results accrue from local assumptions. For example, the plant considered in Section \ref{sec:robot} satisfies local detectability instead of global detectability, and hence Theorem \ref{thm:feedback} guarantees a local asymptotic stability result for the closed-loop system.

It is also easy to show that,
by introducing a closed-loop input $u \coloneqq (u_{1}, u_{2})$ of appropriate dimension and considering the feedback interconnection produced by the conditions
\begin{subequations}
\label{eq:negative-feedback-with-input}
\begin{align}
   u_{c} &= u_{1} - y_{p}, \\
   u_{p} &= u_{2} + y_{c},
\end{align}
the channel $(u_{1}, u_{2}) \mapsto (y_{c}, y_{p})$ is passive with the storage function being the function $V$ from the proof of Theorem \ref{thm:feedback}.
\end{subequations}

\section{Illustrations}
\label{section:Illustrations}

\subsection{TORA}

\begin{figure}
    \centering
    \includegraphics[width=1\linewidth]{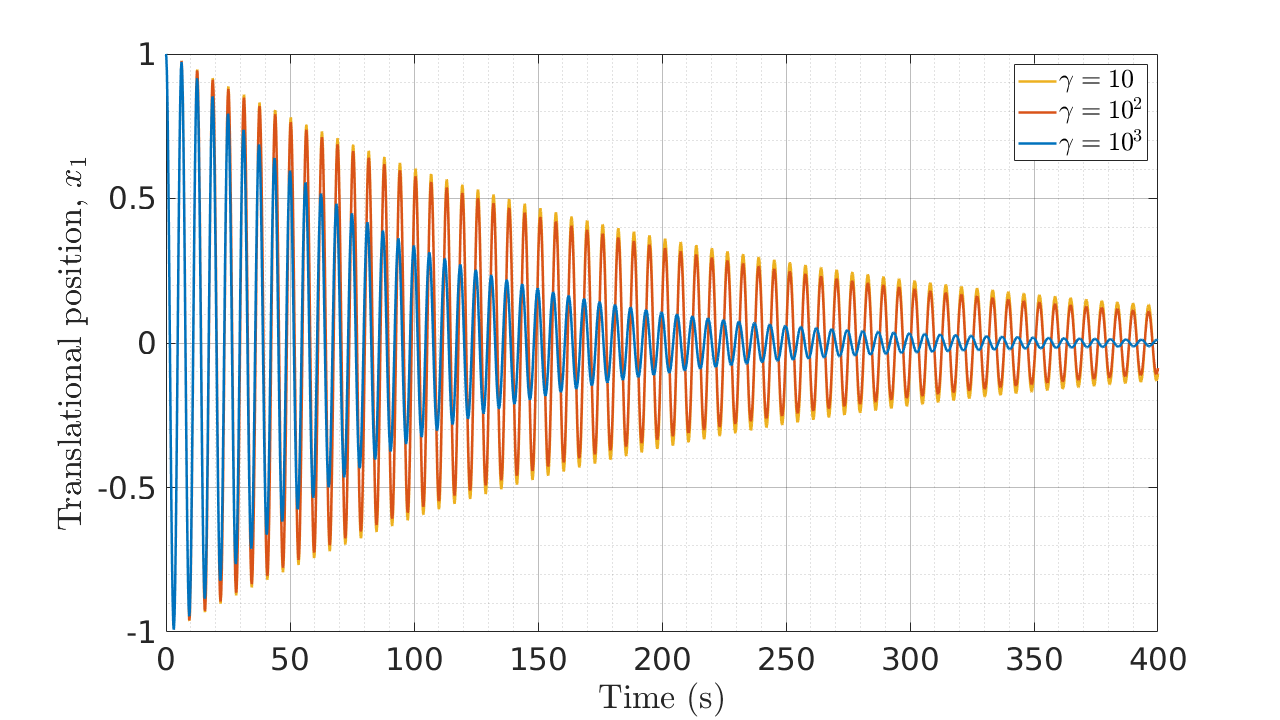}
    \caption{Evolution of the translational position of the TORA system ($\sigma=0.1$) controlled by an unstable FORE with soft resets and various values of the soft-reset parameter $\gamma$. With $\gamma=0$ the origin is not asymptotically stable. With $\gamma>0$ large enough, the origin is asymptotically stable; oscillations are smaller and the convergence is faster for larger $\gamma$.}
    \label{fig:unstable}
\end{figure}

\begin{figure}
    \centering
    \includegraphics[width=1\linewidth]{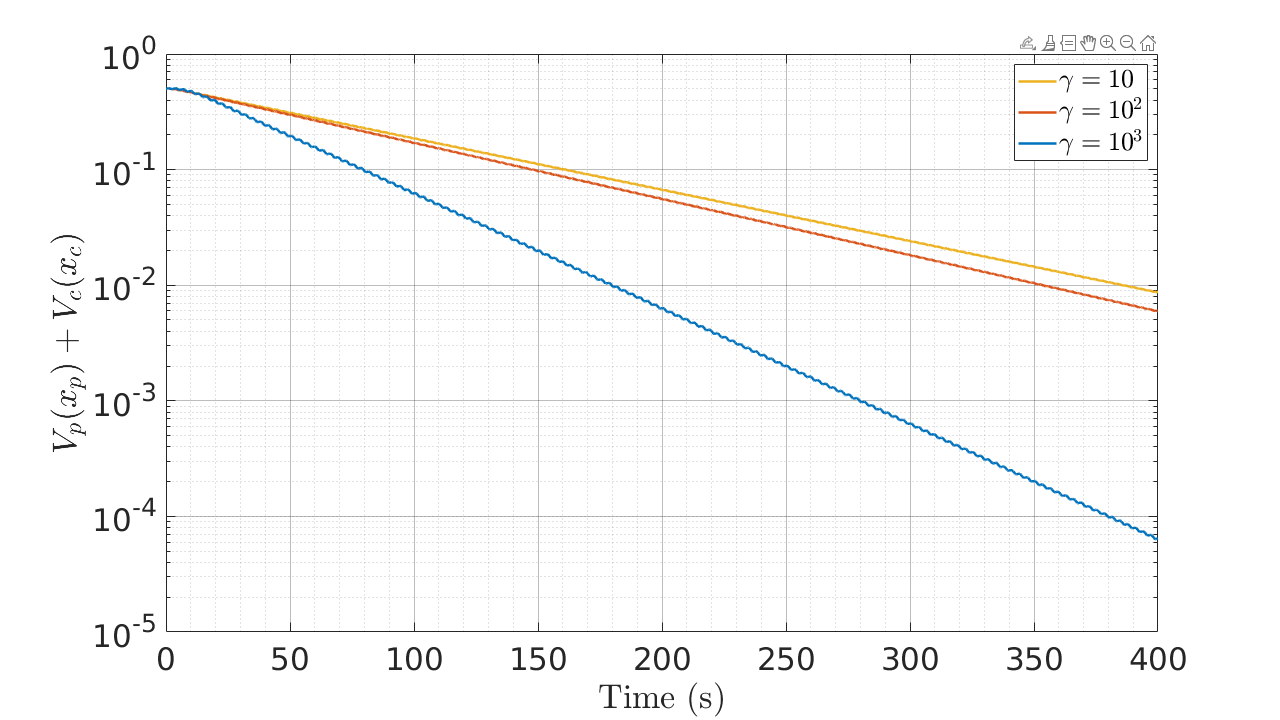}
    \caption{Evolution of the total energy for the TORA system
    ($\sigma=0.1$) controlled by an unstable FORE with soft resets and various values of the soft-reset parameter $\gamma$. The convergence rate increases as $\gamma$ increases.}
    \label{fig:unstable-energy}
\end{figure}

We consider the translational oscillator with rotating actuator (TORA)
from \cite{Wan96} and \cite{Jankovic96}.  The dynamics have the form
\begin{align}
  \left[ \begin{array}{cc} 1 & \sigma \cos(\theta) \\
  \sigma \cos(\theta) & 1
          \end{array}
  \right]
   \left[ \begin{array}{c} \ddot{\theta} \\ \ddot{x} \end{array} \right] = \left[ \! \begin{array}{c} u \\
   -x + \sigma \dot{\theta}^{2} \sin(\theta) 
   \end{array}
   \! \right]
\end{align}
where $\sigma \in (0,1)$, $\theta$ is an angular position, and $x$ is a dimensionless translational position.
   Following \cite[Problem 5.10(b)]{Khalil14}, the preliminary control choice 
$u=-\theta+w$
renders the system passive (in fact, lossless) from $w$ to $\dot{\theta}$, as established by the continuously differentiable, positive definite, radially unbounded energy function
\begin{align}
\label{eq:V-TORA}
 V_{p}(\theta,\dot{\theta},x,\dot{x})  : = & \frac{1}{2} \left( x^{2} + \theta^{2} \right) + \\
 & \frac{1}{2} \left[ \begin{array}{c} \dot{\theta} \\ \dot{x} \end{array} \right]^{T}  \left[ \begin{array}{cc} 1 & \sigma \cos(\theta) \\
  \sigma \cos(\theta) & 1
          \end{array}
  \right]
  \left[ \begin{array}{c} \dot{\theta} \\ \dot{x} \end{array} \right] \nonumber .
\end{align}
Figures \ref{fig:unstable} and \ref{fig:unstable-energy} show the performance of a soft-reset controller using the system data from Example \ref{example}, with $a_{c} = b_{c} = c_{c} = 1$, $d_{c} = 0.01$, $r_{c} = p_{c} = 0$, $\kappa = 1/4$, $\rho = 10^{-3}$, $\varepsilon = 10^{-2}$, and $M = M_{0} + \varepsilon I$ controlling the TORA with $\sigma=0.1$.
The controller is an unstable FORE implemented with soft resets and having the energy function $V_{c}(x) = 0.5|x_{c}|^{2}$. We do not simulate the behavior for $\gamma=0$ since it is immediate that the time derivative
of
\begin{align*}
    t \mapsto V_p(x_{p}(t))+V_{c}(x_{c}(t)) 
\end{align*}
is $2\kappa a_{c} x_{p}^{2}(t) \geq 0$ for all $t \geq 0$ in this case, so that asymptotic stability of the origin is impossible.  The performance for sufficiently large, positive values of $\gamma$ is shown.  The oscillations in the translational position decrease and the convergence rate increases as $\gamma$ increases.



\subsection{Multi-link robotic manipulator}
\label{sec:robot}

\begin{figure}
    \centering
    \includegraphics[width=1\linewidth]{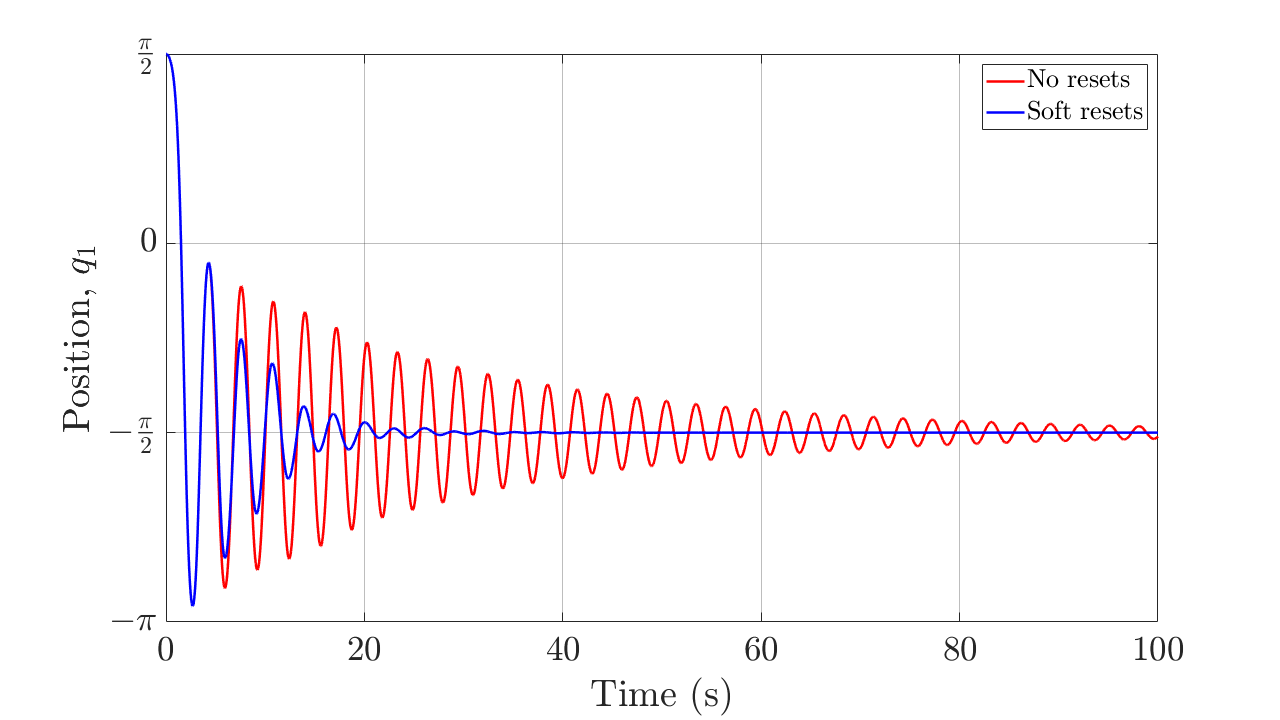}
    \caption{Angular position trajectory of the first link in a $2$-link robotic manipulator, controlled using a stable FORE with angular velocities as inputs and using soft resets.}
    \label{fig:manipulator}
\end{figure}

\begin{figure}
    \centering
    \includegraphics[width=1\linewidth]{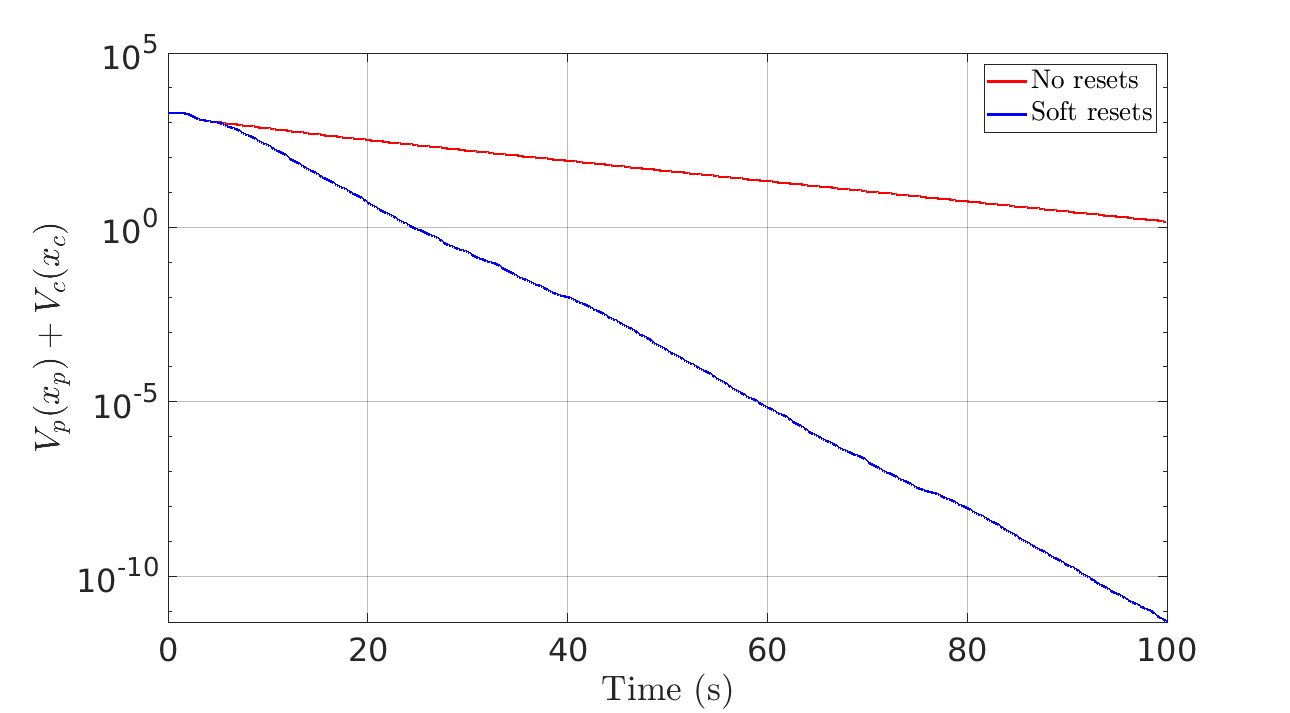}
    \caption{Evolution of the total energy of a $2$-link robotic manipulator, controlled using a stable FORE with angular velocities as inputs and using soft resets.}
    \label{fig:energy_manipulator}
\end{figure}

We consider the planar $2$-link robotic manipulator from \cite[App. A.10]{Khalil14}. The state variable is $q = (q_{1}, q_{2})$, where $q_{1}$ is the angular position of the first link, measured with respect to the horizontal axis of the plane, and $q_{2}$ is the angular position of the second link, measured with respect to the line segment from the first joint to the second joint. We define the potential energy as
\begin{align*}
    P(q) &\coloneqq 784.8\sin(q_{1}) + 245.25\sin(q_{1} + q_{2}) \\
    &\quad + \frac{\varrho}{2} \left(\left|q_{1} + \frac{\pi}{2}\right|^2 + \left|q_{2}\right|^2\right),
\end{align*} 
where $\varrho$ is a small positive number to ensure radial unboundedness and positive definiteness of $q \mapsto P(q) - P(q^{*})$ with respect to $q^{*} \coloneqq (-\frac{\pi}{2}, 0)$. With $g(q) \coloneqq \partial{P}(q)/\partial{q}$ for all $q \in \Re^{n}$, the dynamic equation of the manipulator is
\begin{align}
    \ddot{q} = M^{-1}(q)[u - C(q, \dot{q})\dot{q} - g(q)],
\end{align}
with $u$ being the system input, and with $M$ and $C$ given by \cite[Eqs. A.36-A.38]{Khalil14}. Defining the state $x_{p} = (q, \dot{q})$, passivity from $u$ to $\dot{q}$ can be shown using the continuously differentiable, positive definite, radially unbounded energy function
\begin{align}
    V_{p}(x_{p}) &= \frac{1}{2}\dot{q}^{T}M(q)\dot{q} + P(q) - P(q^*).
\end{align}
For the system having the translated state $\tilde{x}_{p} \coloneqq (\tilde{q}, \dot{q})$ with $\tilde{q} \coloneqq q - q^{*}$, global detectability does not hold because, for small $\varrho$, $g(q)$ can be zero at some points where $q \neq q^{*}$. However, $g(q)$ is uniquely zero at $q^{*}$ in the region where $q - q^{*} \in [-\pi, \pi]^{2}$. Thus, local detectability holds for trajectories $\tilde{x}_{p}$ for which $\tilde{q}(t) \in [-\pi, \pi]^{2}$ for all $t$. Figure \ref{fig:manipulator} shows the trajectory of $q_{1}$ 
and Figure \ref{fig:energy_manipulator} shows the evolution of the energy when the system is controlled using measurements of $\dot{q}$, with the controller having input $u_{c} = -\dot{q}$ and having the form
\begin{subequations}
\label{eq:first_order}
\begin{align}
    f_{c}(z_{c}) &= A_{c}x_{c} + B_{c}u_{c}, \\ 
    g_{c}(z_{c}) &= 0, \\
    h_{c}(z_{c}) &= C_{c}x_{c},
\end{align}
\end{subequations}
with $A_{c} = -I_{n_{c}}$, $B_{c} = 100I_{n_{c}}$, and $C_{c} = I_{n_{c}}$. With the chosen system matrices, the controller's energy function is given by $V_{c}(x) = 0.005|x_{c}|^{2}$. The initial condition is $(\pi/2, -\pi/4)$, which we have chosen so that, via the closed-loop stability verified by the Lyapunov function $V_{p}(x_{p}) + V_{c}(x_{c})$, the trajectories satisfy $\tilde{q}(t) \in [-\pi,\pi]^{2}$ for all $t$. The soft resets are implemented with $\gamma = 10^{3}$ and
\begin{align}
\label{eq:M_first_order}
    M &= \left[\begin{array}{cc}
        0 & -\frac{1}{2}I_{n_{c}}  \\
        -\frac{1}{2}I_{n_{c}} & 0
    \end{array}\right].
\end{align}

\subsection{Strongly convex, non-quadratic accelerated optimization}
\begin{figure}
    \centering
     \begin{subfigure}[b]{0.45\textwidth}
         \centering
         \includegraphics[width=\linewidth]{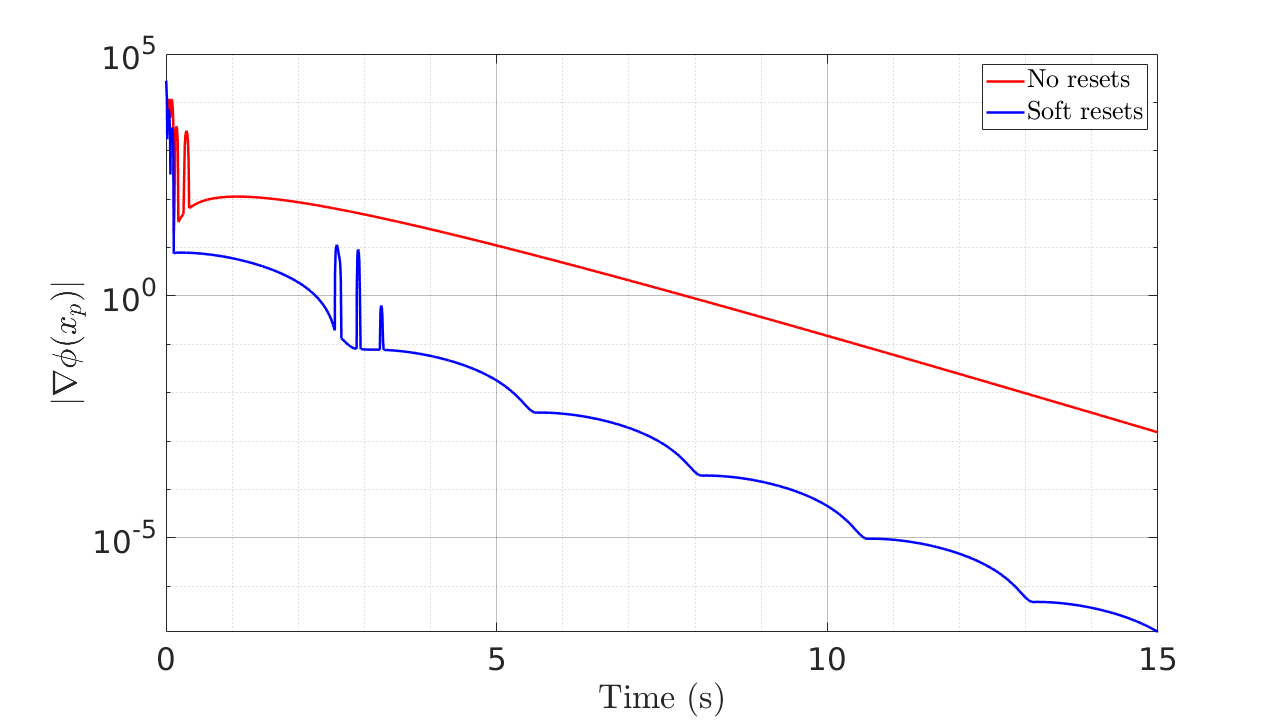}
         \label{fig:sc_2b}
     \end{subfigure}
     \hfill
     \begin{subfigure}[b]{0.45\textwidth}
         \centering
         \includegraphics[width=\linewidth]{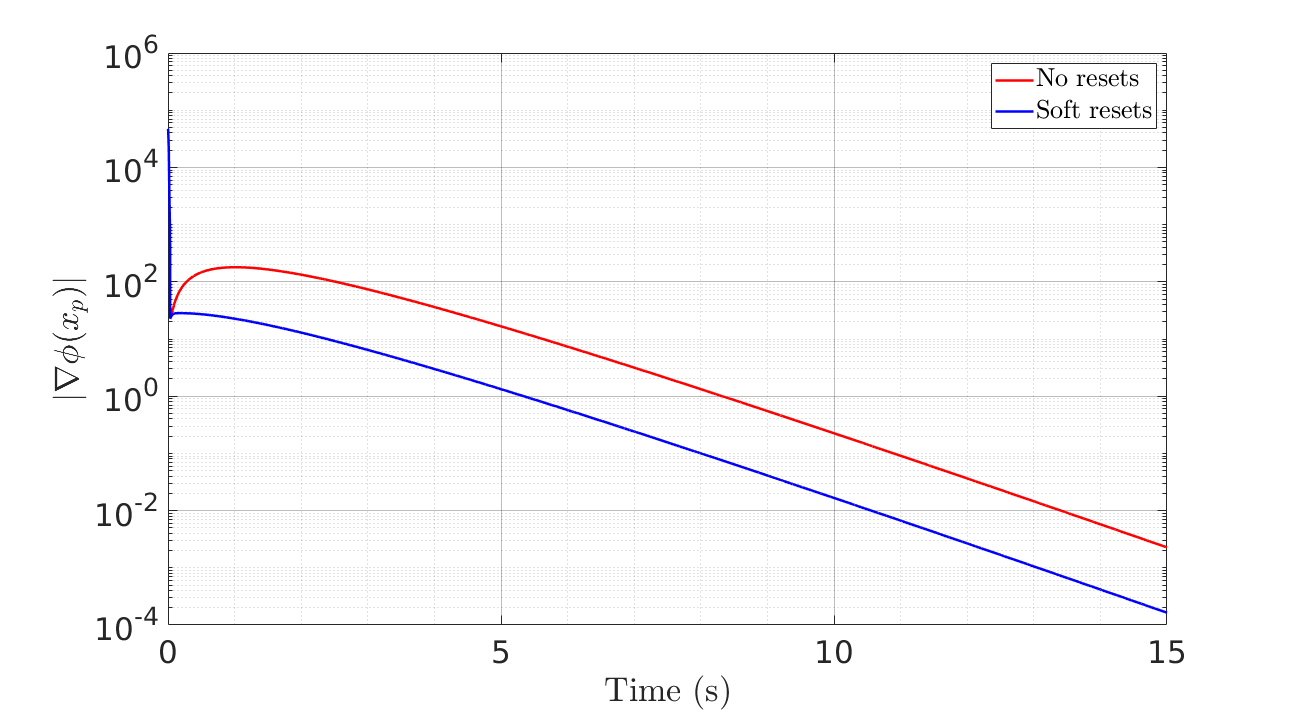}
         \label{fig:sc_5b}
     \end{subfigure}     
        \caption{Performance of reset control for strongly convex optimization.}
        \label{fig:strongly_convex}
\end{figure}

We consider the strongly convex function $\phi: \Re^{n_{p}} \to \Re$ from \cite[Eq. 17]{van-scoy-2018}, given by
\begin{align}
     \phi(x) &= \sum_{i=1}^{p} \hat{\phi}(a_{i}^{T}x - b_{i}) + \frac{1}{2}|x|^2, \\
     \hat{\phi}(\alpha) &= \begin{dcases}
        \frac{1}{2}\alpha^{2} \exp(-r/\alpha), & \alpha > 0 \\
        0 & \alpha \leq 0,
    \end{dcases} 
\end{align}
with $r = 10^{-6}$. We randomly generate the entries of $A$ and $b$ as described in \cite{van-scoy-2018}, such that $|A| = \sqrt{L - 1}$, where $L = 10^{4}$ is the Lipschitz constant of $\nabla\phi$. To apply our proposed reset control approach toward the minimization of $\phi$, we define the plant system
\begin{align}
\label{eq:sc_plant}
    f_{p}(x_{p}, u_{p}) &= u_{p}, \\ 
    h_{p}(x_{p}, u_{p}) &= \nabla\phi(x_{p}),
\end{align}
and we take the control system to be defined by \eqref{eq:first_order} with $n_{c} = n_{p}$, $A_{c} = -KI_{n_{c}}$, $B_{c} = I_{n_{c}}$, and $C_{c} = I_{n_{c}}$, where $K \in \Re_{>0}$ is a tuning parameter. The top plot in Figure \ref{fig:strongly_convex} shows the evolution of $|\nabla\phi(x_{p})|$ for the case of $n_{p} = 5$, $p = 10$, and $K = 2$. The soft resets are implemented with $K = 1$, $\gamma = 30$, and $M$ given by \eqref{eq:M_first_order}. The bottom plot in Figure \ref{fig:strongly_convex} shows the evolution of $|\nabla\phi(x_{p})|$ for the case of $n_{p} = 50$, $p = 5$, and $K = 2$. The soft resets are implemented with $K = 2$, $\gamma = 20$, and $M$ given by \eqref{eq:M_first_order}.
    
\section{Conclusions}

We have introduced nonlinear, passive, soft-reset control systems as an alternative to hard-reset control systems.  Soft-reset control systems avoid the formalism of hybrid systems and are straightforward to implement, without any temporal regularization of other ``robustifying'' mechanisms.  A passive hard-reset control system can be implemented as a soft-reset control system when it admits a strongly convex energy function.  Asymptotic stability for the origin of the interconnection of a passive soft-reset control system with a nonlinear, passive plant has been established, under weak detectability conditions.  The theory has been illustrated on two mechanical systems and also in the context of optimization of a strongly convex, non-quadratic objective function.

\bibliographystyle{unsrt}
\bibliography{refs}

\end{document}